\documentclass[prb,a4paper,twocolumn,showpacs,superscriptaddress]{revtex4-1}

\usepackage[utf8x]{inputenc}
\usepackage{lmodern}

\usepackage{amssymb}
\usepackage{amsmath}
\usepackage{amsbsy}
\usepackage{bm}
\usepackage{graphicx}
\usepackage{color}
    \newcommand{\I}{\mathrm{i}}

\begin{document}

\title{Orbital polaron in double exchange ferromagnets}

\author{T. L. van den Berg}
\author{P. Lombardo}
\affiliation{Aix-Marseille Université, IM2NP-CNRS UMR 7334, Campus St. Jérôme, Case 142, 13397 Marseille, France}

\author{R. O. Kuzian}
\affiliation{Institute for Problems of Materials Science, NASU, Krzhizhanovskogo 3, 03180 Kiev, Ukraine}
\affiliation{Donostia International Physics Center (DIPC), ES-20018 Donostia-San Sebastian, Spain}

\author{R. Hayn}
\affiliation{Aix-Marseille Université, IM2NP-CNRS UMR 7334, Campus St. Jérôme, Case 142, 13397 Marseille, France}

\date{\today}

\begin{abstract}
We investigate the spectral properties of the two-orbital Hubbard model, including the pair hopping term, by means of the dynamical mean field method. This Hamiltonian describes materials in which ferromagnetism is realized by the double exchange mechanism, as for instance manganites, nickelates or diluted magnetic semiconductors. The spectral function of the unoccupied states is characterized by a specific equidistant three peak structure. We emphasize the importance of the double hopping term on the spectral properties. We show the existence of a ferromagnetic phase due to electron doping near \(n=1\) by the double exchange mechanism. A quasi-particle excitation at the Fermi energy is found that we attribute to what we will call an orbital polaron. We derive an effective spin-pseudospin Hamiltonian for the two-orbital double exchange model at \(n=1\) filling to explain the existence and dynamics of this quasi-particle.
\end{abstract}


\maketitle

%%%%%%%%%%%%%
\section{Introduction}
\label{sec.introduction}
%%%%%%%%%%%%%
The double exchange mechanism for ferromagnetism in magnetic perovskite compounds was proposed in 
1951 by Zener, \cite{Zener-1951xz} and further developed by de Gennes. \cite{Gennes-1960ly}
It is a mechanism that works for transition metal ions with a partially filled $d$-shell building up a local spin. 
Additional charge carriers in the $d$-orbitals that move through the crystal prefer a ferromagnetic arrangement 
with the local spin according to Hund's first rule and induce in this way a ferromagnetic
long range order. This mechanism is the origin of ferromagnetic behavior in manganites which became 
of technological  importance after the discovery of giant magnetoresistance (GMR) in these compounds.
This discovery led to a renew of interest in the double exchange mechanism.\cite{Brink-1999vn}

Manganites are only one class in the fascinating world of transition metal compounds that also include
nickelates, cuprate superconductors and its parent compounds, and the recently discovered iron 
pnictide superconductors. The common feature to all these compounds is the strong correlation
in the $d$-shell which leads to bad metal behavior, magnetism, spectral weight transfer, superconductivity, 
orbital excitations and other phenomena that are far from being well understood. To clarify the role 
of orbital fluctuations (that are absent in cuprate superconductors) we consider here the minimal model with two orbital degrees of freedom per site. We consider the region of strong electron correlations, i.e. large values of the on-site Coulomb repulsion $U$. The system is insulating at quarter filling ($n=1$) and we will show that additional electrons ($n>1$) lead to ferromagnetism by the double exchange
mechanism that we investigate in detail.

Other materials where the double exchange mechanism may appear are diluted magnetic 
semiconductors (DMS), a prominent example is Mn-doped GaAs. In this system, however, the kinetic $p$-$d$-exchange mechanism, is generally believed to be more important.\cite{Zener-1951ct} This mechanism, also, proposed by Zener in the same year (1951), demands charge carriers (usually holes) in the $p$-orbitals.  Although one may expect the kinetic \(p\)-\(d\) exchange mechanism in GaAs:Mn, in GaN:Mn the double exchange mechanism is likely to be more important. The reason is the localization of the Mn \(3d\) orbital which is much more pronounced in GaN:Mn than in GaAs:Mn.\cite{Krstajic-2004kx} Despite the considerable interest and effort put into unraveling the interaction mechanisms in DMSs, no clear criteria to distinguish between both Zener's mechanisms is known. In the following, we analyze the double exchange mechanism and propose a characteristic that might be used to distinguish it from other mechanisms. We report the discovery of a narrow quasi-particle peak at the Fermi level with a spin parallel to the ferromagnetic order. 

Narrow quasiparticle peaks corresponding to large effective masses are already well known in the physics 
of strongly correlated electrons. The most prominent examples are the Kondo peak of heavy fermions and
the spin polaron in the $t$-$J$ model for cuprate superconductors.~\cite{Martinez-1991nx,Eder-1990fk,Kane-1989kx,Mishchenko-2001fk} In both cases, there are spin fluctuations
coupled to the moving charge leading to the heavy mass, and in both cases, the spin correlation between 
the local magnetic moment and the itinerant charge carrier is of antiferromagnetic nature. We will show that 
both aspects are different for quasiparticles in double exchange ferromagnets: the electrons are coupled 
to orbital fluctuations and the arrangement between local spin and itinerant electron is ferromagnetic. In our 
case, spin fluctuations are excluded since the system is close to saturated ferromagnetism ($n=1$ and \(|M|=1\)). But 
at quarter filling, the system shows alternating orbital order. Therefore, there exists a close analogy between a 
hole moving in the antiferromagnetic background of cuprate superconductors 
(spin polaron)\cite{Martinez-1991nx,Eder-1990fk,Kane-1989kx,Mishchenko-2001fk} and an electron moving in the alternating orbital environment of double exchange ferromagnets. We will use this analogy to present a concise interpretation of the quasi-particle that we will call an orbital polaron. The term "orbital polaron" was originally introduced by R. Kilian and G. Khaliullin for a quasi-particle for which the charge degree of freedom is not only coupled to orbital fluctuations, but also to the lattice.~\cite{Kilian-1999uq} Later the term was used in studies on effective, low-energy $t$-$J$ like Hamiltonians in the field of manganites for orbital quasi-particles, but without coupling to the lattice.~\cite{Brink-2000kl,Daghofer-2004oq,Wohlfeld-2008ve,Wohlfeld-2009qf} We re-discover the orbital polaron here starting from the complete two-orbital Hubbard Hamiltonian with the
pair-hopping term.

The double exchange mechanism in the two-orbital Hubbard model was already studied before. Many works concentrated on the case \(n=1\). Below, we use the dynamical mean field theory (DMFT) with the non-crossing approximation (NCA) as an impurity solver. The phase diagram in DMFT was already calculated in Refs. \onlinecite{Peters-2010fk} and  \onlinecite{Peters-2011uq}, however, the paired hopping between different orbitals was neglected. On the other hand, the spectral weight transfer of the complete Hamiltonian was analyzed in an analytical way and several characteristic features were found that we find back using the DMFT approach.\cite{Lee-2011uq} Here, we include the pair-hopping term (or double hopping term) into our study and we concentrate on the spectral function in the slightly doped case ($n>1$). We will show that the pair-hopping term has indeed only small influence on the phase diagram, but it is very important for the spectral function. It leads to characteristic features in the unoccupied parts of the spectral function confirming the analytical results of Lee and Phillips.\cite{ Lee-2011uq}

The paper is set up as follows. After introducing the double exchange Hubbard model and the 
numerical method (Secs. \ref{sec.model} and \ref{sec.numerical_method}), we present the magnetization diagram in Sec.~\ref{sec.spectral}. Special care is taken to analyze the spectral properties and the importance of the pair-hopping term is emphasized. In Sec.~\ref{subsec.QPpeak} the quasi-particle peak at the Fermi level is investigated. To understand the model at quarter filling ($n=1$) we derive 
an effective low-energy model in the large $U$-case in Sec.~\ref{sec.Heff} and we show that it leads to a ferromagnetic
phase with alternating orbital order. That observation allows a concise interpretation of 
the sharp quasi-particle peak in terms of an orbital polaron (Sec.~\ref{sec.orbital_polaron}). The paper is closed with a discussion and conclusion (Sec.~\ref{sec.conclusion}).

%%%%%%%%%%%%%%%%%%%%%%
\section{Double exchange Hubbard model}
\label{sec.model}
%%%%%%%%%%%%%%%%%%%

We will write the double exchange Hamiltonian as \( \hat{H}  =  \hat{H}_t + \hat{H}_d\), with
\begin{equation}
\hat{H}_t  =  t\sum_{r,g,\alpha,\sigma}d_{r,\alpha,\sigma}^{\dagger}d_{r+g,\alpha,\sigma},\label{eq.Hkin}\\
\end{equation}
and
\begin{align}
\hat{H}_{d} & =  U\sum_{r,\alpha}\hat{n}_{r,\alpha,\uparrow}\hat{n}_{r,\alpha,\downarrow}-2J\sum_{r}\hat{\bm{S}}_{r,a}\hat{\bm{S}}_{r,b} \label{eq.HDE1} \\
 & +  (U^{\prime}-\frac{J}{2})\sum_{r,\sigma,\sigma^{\prime}}\hat{n}_{r,a,\sigma}\hat{n}_{r,b,\sigma^{\prime}} \label{eq.HDE2} \\
 & +  J^{\prime}\sum_{r,\alpha\neq\beta}d_{r,\alpha,\uparrow}^{\dagger}d_{r,\alpha,\downarrow}^{\dagger}d_{r,\beta,\downarrow}d_{r,\beta,\uparrow} \label{eq.Hd}
\end{align} 
where \(r\) is the site index, \(\alpha,\beta=a,b\) is the orbital number,
\(\sigma=\uparrow,\downarrow\) is the spin projection, \(g\) is a vector
that joins nearest neighbors. The operators \(\hat{\bm{S}}_{r,\alpha}= (S_x,S_y,S_z)_{r,\alpha}\) 
are the spins of an electron on site \(r\) in orbital \(\alpha\). To find eigenstates of the spin term 
it is convenient to rewrite it \(\hat{\bm{S}}_{r,a} \cdot \hat{\bm{S}}_{r,b} = \frac12 \left(S^+_{r,a} S^-_{r,b} +  S^-_{r,a} S^+_{r,b} \right) + S^z_{r,a} S^z_{r,b} \).
The Hamiltonian (\ref{eq.HDE1})-(\ref{eq.Hd}) describes Coulomb interaction \(\hat{H}_{d}\) in Kanamori's simplified form,
where the parameter \(U\) denotes the repulsion of electrons occupying the same orbital, \(U^{\prime}=U-2J\) is the repulsion of electrons
occupying different orbitals, \(J\) is the Hund exchange term, \(J^{\prime}=J\) is the double hopping term, which is always present when exchange
interactions take place. Here we retain different notations for the parameters \(J,J^{\prime}\) in order to clarify the origin of different terms, and to facilitate the comparison with the results of some papers (e.g. Ref.\onlinecite{Kuei-1997fk}) where   \(J^{\prime}\) is not taken into account. The term \(\hat{H}_t\) is the kinetic energy. For simplicity we only introduced hopping terms that are diagonal in orbital and spin indices and we consider a semi-circular density of states (DOS) being realized on the Bethe lattice. The band width is \(W=4t=2\) and we chose the half-bandwidth as the unit of energy, which is of the order of the electron volt. We take the Coulomb repulsion energy \(U=4W=8\) and the Hund exchange term \(0.5<J<2\), which is a relatively small parameter.

To solve this Hamiltonian numerically we use dynamical mean field theory with the non-crossing approximation. The following paragraph gives a quick introduction to this method.

%%%%%%
\section{Numerical method}
\label{sec.numerical_method}
%%%% %%
In the framework of the dynamical mean field theory (DMFT)~\cite{Georges-1996fk}, all fermionic degrees of freedom
except those for a central site $r=0$ are integrated out, 
and the double exchange Hamiltonian described above
can be mapped onto an effective single impurity Anderson model (SIAM)
described by an effective Hamiltonian $ \hat{H}^{\mathrm{eff}}$~:
\begin{equation}
   \hat{H}^{\mathrm{eff}}=\hat{H}_{d}^{\mathrm{eff}}+\hat{H}_{t}^{\mathrm{eff}}~, 
\end{equation}
where
\begin{align}
\hat{H}_{d}^{\mathrm{eff}} & =  U\sum_{\alpha}\hat{n}_{0,\alpha,\uparrow}\hat{n}_{0,\alpha,\downarrow}-2J\hat{\bm{S}}_{0,a}\hat{\bm{S}}_{0,b} \nonumber \\
 & +  (U^{\prime}-\frac{J}{2})\sum_{\sigma,\sigma^{\prime}}\hat{n}_{0,a,\sigma}\hat{n}_{0,b,\sigma^{\prime}} \\
 & +  J^{\prime} \sum_{\alpha\neq\beta}d_{0,\alpha,\uparrow}^{\dagger}d_{0,\alpha,\downarrow}^{\dagger}d_{0,\beta,\downarrow}d_{0,\beta,\uparrow}  \nonumber
\end{align}
refers to the impurity, which is coupled to the effective medium
\begin{align}
   \hat{H}_{t}^{\mathrm{eff}}=&\sum_{{\vec k}, \alpha,\sigma}
   \left(V_{\alpha,\vec k } c^{\dagger}_{\alpha,{\vec k}, \sigma} d_{0,\alpha,\sigma}+H.c.\right) \nonumber\\
   &+\sum_{{\vec k} ,\alpha,\sigma}\varepsilon_{\alpha,\vec k, \sigma} c^{\dagger}_{\alpha,{\vec k}, \sigma}c_{\alpha,{\vec k}, \sigma}~.
\end{align}
$V_{\vec k,\alpha}$ represents the hybridization between the site $r=0$ and the
effective medium corresponding to orbital $\alpha$. $\varepsilon_{\alpha,{\vec k},\sigma}$ is
the band energy of the corresponding effective medium. $c^{\dagger}_{\alpha,{\vec k}, \sigma}$
(respectively $c_{\alpha,{\vec k}, \sigma}$) is the creation (respectively
annihilation) operator of an electron in the effective medium $\alpha$.
We have therefore to determine two spin-dependent self-consistent 
effective baths, as in Ref.~\onlinecite{Lombardo-2002uq}, but here we treat the full double exchange Hamiltonian.
Moreover, in the present study we go beyond the study of the paramagnetic state, 
since we are interested in the stability of the spontaneous ferromagnetic phase.
The effective medium is characterized by the effective dynamical hybridization~:
\begin{equation}
   {\cal J}_{\alpha \sigma} (\omega)=\sum_{{\vec k}}\frac{|V_{\alpha,\vec k}|^2}{\omega+i0^+-\varepsilon_{\alpha, \vec k, \sigma}}~.
\end{equation}
The equation of motion for the Hamiltonian $ \hat{H}^{\mathrm{eff}}$
gives the following equation for the retarded Green's function~:
\begin{equation}
   G_{\alpha\sigma}(\omega)^{-1}=\omega-\Sigma_{\alpha\sigma}(\omega)
   -{\cal J}_{\alpha\sigma} (\omega)~.
\end{equation}
Comparing this equation to the following property that a Green's function satisfies on a Bethe lattice~:
\begin{equation}
   G_{\alpha\sigma}(\omega)^{-1}=\omega-\Sigma_{\alpha\sigma}(\omega)
   -t^2 G_{\alpha\sigma}(\omega)~,
\end{equation}
we get the self-consistent equations of the DMFT~:
\begin{equation}
{\cal J}_{\alpha \sigma} (\omega)=t^2 G_{\alpha\sigma}(\omega)~.
\end{equation}
As an impurity solver for the DMFT, we used the noncrossing approximation (NCA)~\cite{Bickers-1987kx,Pruschke-1993vn}.
Despite the known pathology of the NCA at very low temperature when applied to the SIAM, it gives reliable results  for temperatures
down to a fraction of the Kondo temperature. As a self-consistent and conserving approximation, it also displays the correct scaling
behavior and reproduces the relevant energy scales. In the DMFT framework the NCA has been applied successfully to various compounds like cerium~\cite{Zolfl-2001ys} , La$_{1-x}$Sr$_{x}$TiO$_{3}$~\cite{Zolfl-2000ly}and SrRuO$_{3}$~\cite{Anisimov-2002zr} allowing the first prediction concerning the orbitally selective Mott transition.

Within the NCA, propagators $P_m(\omega)$ and self-energies $\Sigma_m(\omega)$ for the $16$ local eigenstates of the local Hamiltonian   $\hat{H}_{d}^{\mathrm{eff}}$   are introduced. The sixteen coupled integral equations between local propagators and self-energies are solved numerically for each effective medium until convergence.

%%%%%%%%%%%%%%%%%%%
\section{Spectral properties}
\label{sec.spectral}
%%%%%%%%%%%%%%%%%%%
%FIGURE Magnetization diagram
\begin{figure}[!t]
\centering%
\includegraphics[width=1.0\linewidth]{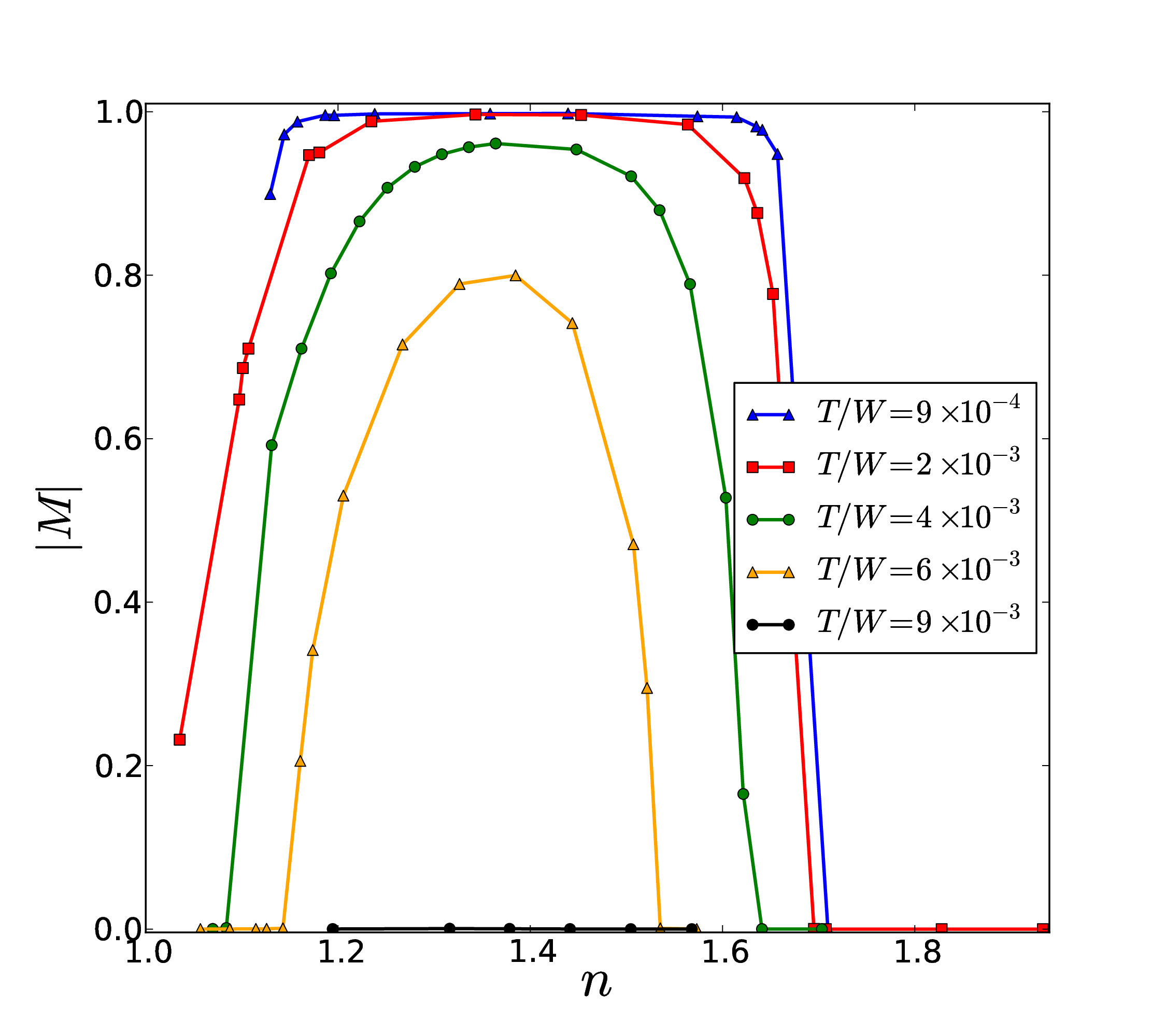}
\caption{\label{fig.phasediag_J125}Magnetization as a function of filling \(n\) for different temperatures (see legend), here \(J=1.25\). We remind the reader that the half-bandwidth is the energy unit. The ferromagnetic phase is centered around \(n=1.4\), at low temperatures it approaches \(n=1\).}
\end{figure}
The phase diagram of this model was studied quite extensively by Peters \textit{et al}. \cite{Peters-2010fk,Peters-2011uq} The phase diagrams obtained by our method are in good agreement with their results. The magnetization diagram in Fig.~\ref{fig.phasediag_J125} is typical for all magnetization diagrams for \(0.75<J<2.0\). At \(J=0.5\) there is no ferromagnetic phase at all, and at \(J=0.75\) its extension is small and centered around \(n\approx1.4\). The domain of ferromagnetism increases with \(J\) and for \(J \gtrsim 1.5\) the ferromagnetic phase includes quarter filling.\cite{Peters-2010fk} 

Since we did not introduce two sublattices, our numerical NCA procedure is unable to distinguish between orbitally ordered and orbital liquid phases in the ferromagnetic domain. However, as will be explained in Sec.~\ref{sec.Heff}, there are strong reasons to expect a saturated ferromagnetic phase with alternating orbital order near \(n=1\) at low temperatures (see also Ref.~\onlinecite{Peters-2010fk}). For \(n=1+x\) electron doping will destroy the orbital order (see Sec.~\ref{sec.orbital_polaron}) in close analogy to the destruction of antiferromagnetic long range order by hole doping in cuprate compounds. By the same analogy we expect a very small critical value of electron doping of only a few percent. Summarizing, we expect three different phases at zero temperature for \(n\) varying between \(1\) and \(2\) (and taking \(J=1.25\) as a representative value as in Fig.~\ref{fig.phasediag_J125}): \((i)\) The ferromagnetic and orbitally ordered phase close to \(n=1\), \((ii)\) a saturated ferromagnetic and orbital liquid phase at higher values of electron doping, and \((iii)\) for \(n>1.7\) one gets a paramagnetic phase without orbital order.

%FIGURE Comparison Lee&Phil
\begin{figure}
\centering%
\includegraphics[width=1.0\linewidth]{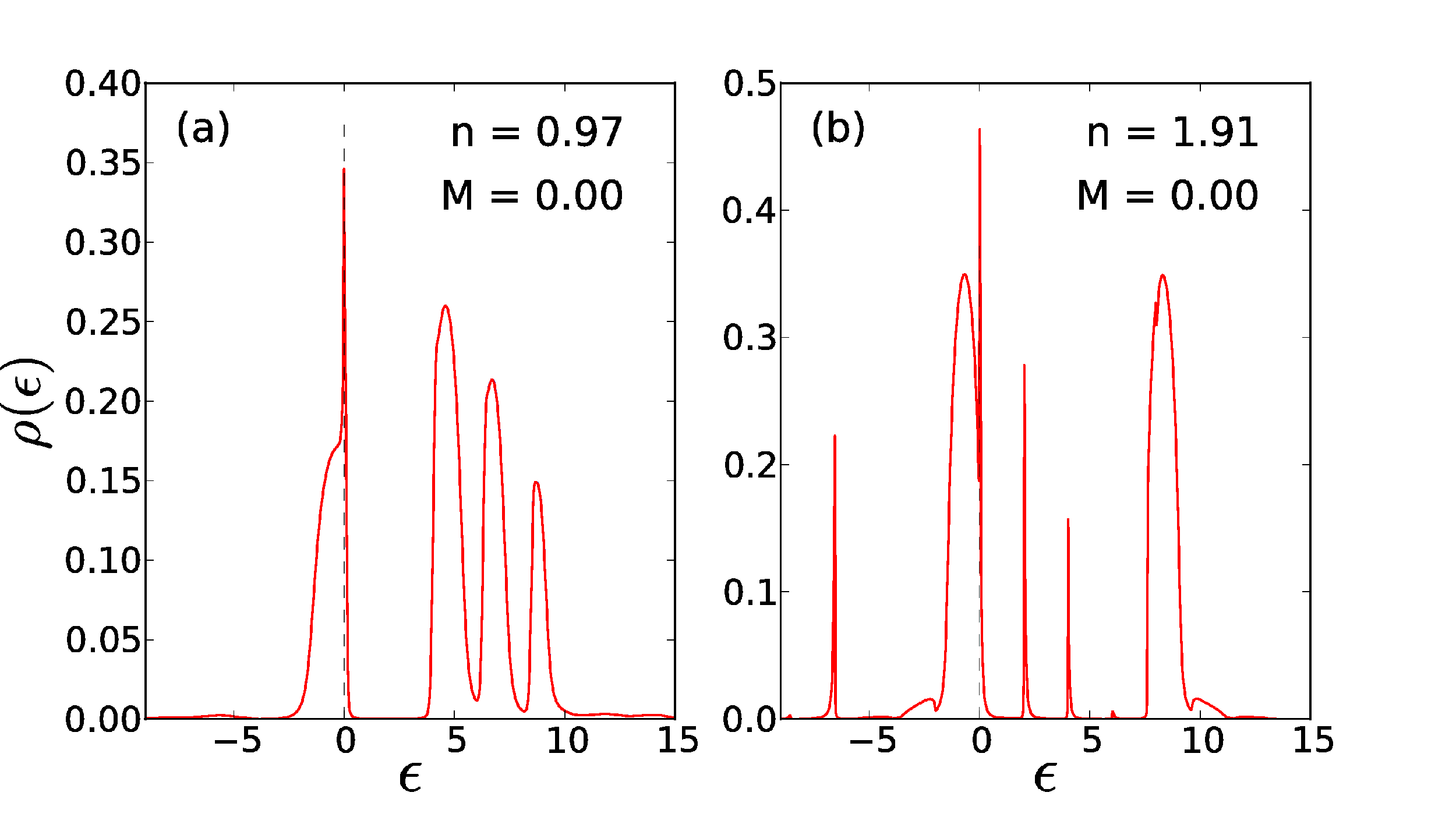}
\caption{\label{fig.comparisonLee}Spectral function for \(n=1-x\) and \(n=2-x\), at temperature \(T=4.3\times10^{-3}\) and \(J=1.0\). The origin of the energy axis corresponds to the Fermi energy. }
\end{figure}

Figure ~\ref{fig.comparisonLee} shows the spectral density 
\(\rho (\epsilon ) = -\frac1\pi \textrm{Im}\{G(\epsilon +\I 0)\}\). For \(n=1\pm x\) or \(n=2\pm x\) we can compare our results to those obtained in a recent paper by W.-C. Lee and P. W. Philips.\cite{Lee-2011uq} Around \(n=1\) one clearly finds the series of three consecutive upper Hubbard bands, with the spectral weight distribution 3:2:1 (see Fig.~\ref{fig.comparisonLee}). 
We will see in Sec.~\ref{subsec.doublehopping} that this spectral weight repartition is due to the double hopping term. Without explicit consideration of this term the three-peak structure of the unoccupied states is absent, leading to qualitatively different behavior. Close to \(n=2\) only two bands remain, each carrying the same spectral weight. The spectral weight transfer we observe between \(n=1-x\) and \(n=2+x\) is also in good agreement with earlier published analytic results.\cite{Lee-2011uq} The narrow peaks at the Fermi energy we observe in Fig.~\ref{fig.comparisonLee} are both quasi-particle peaks, we will discuss this in Sec.~\ref{subsec.QPpeak}.

%%%%%%%%%%%%
\subsection{Dependence on $J$}
\label{subsec.Jdependence}
\begin{figure}
\centering%
\includegraphics[width=\linewidth]{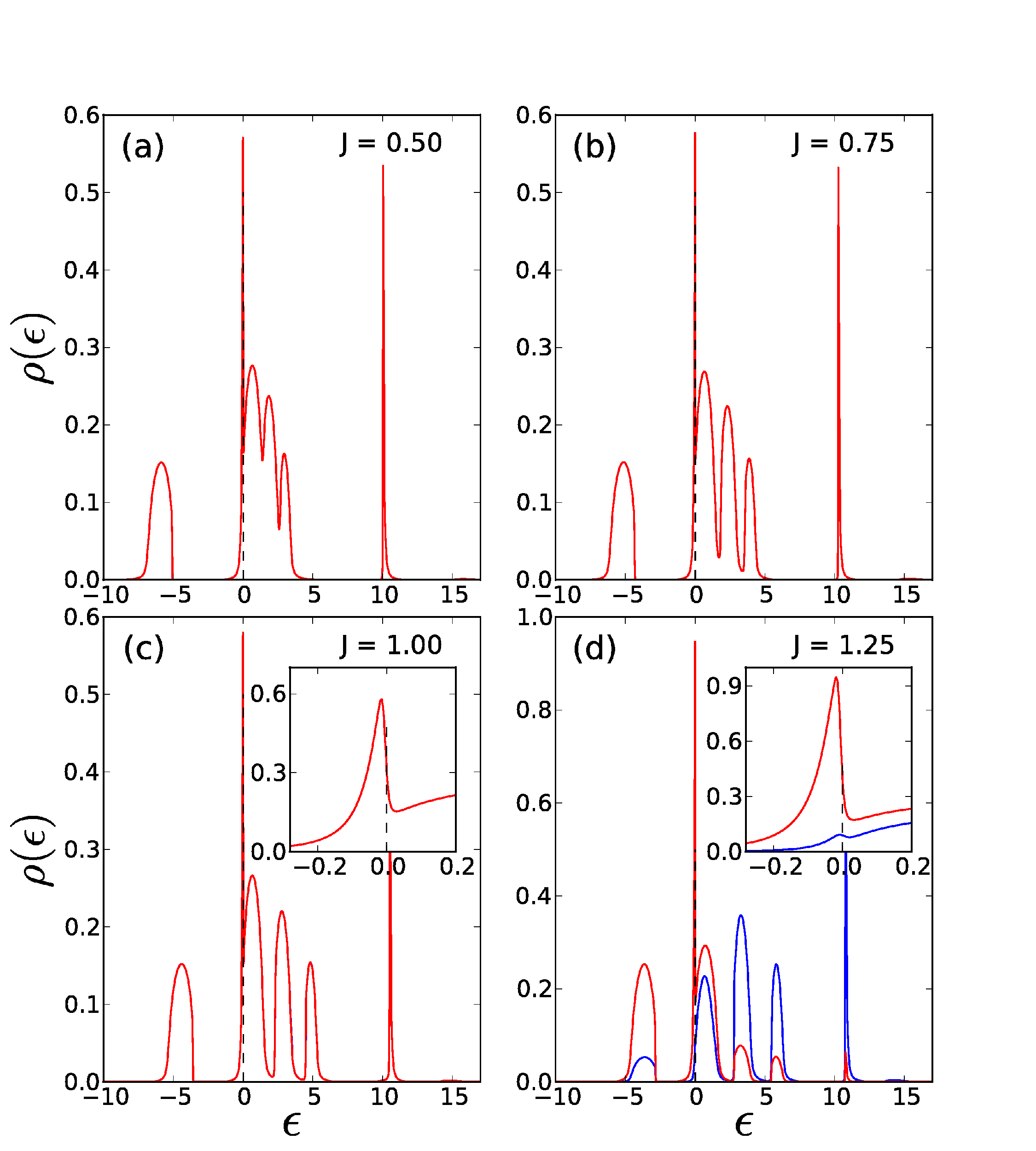}
\caption{\label{fig.QPpeak_Jdep}Spectral function of the two spin states (red and blue) for different values of Hund's exchange \(J\) for \(n=1.10\), \(T=4.3\times 10^{-3}\) and \(J=0.5-1.25\). (a) For \(J=0.5\) no magnetic phase is observed. (b)-(c) For \(J=0.75\) and  \(J=1.0\) the magnetic phase does not include \(n=1.10\). (d) For \(J=1.25\) the ferromagnetic phase is close to \(n=1\), here \(M=0.68\). The quasi-particle peak (inset) is discussed in Sec.~\ref{subsec.QPpeak}.}
\end{figure}

%%%%%%%%%%%%
It is instructive to study the $J$ dependence of the spectral function (Fig.~\ref{fig.QPpeak_Jdep}). Near \(n=1\)  (\(n=1.10\) for Fig.~\ref{fig.QPpeak_Jdep}) the unoccupied part of the spectral function of the two-orbital 
Hubbard model is very characteristic with a specific equidistant three peak structure.
%FIGURE J dependence 
The energy difference between the peaks is approximately \(2 J\) which leads to a better separation of the peaks with the increase of \(J\). The energy difference between the lower Hubbard band and the first upper band is \(U-3J\).\cite{Lee-2011uq} As \(J\) increases the tendency towards ferromagnetism increases and is realized in  Fig.~\ref{fig.QPpeak_Jdep} (d) for $J=1.25$. In the ferromagnetic case, the weight distribution 3:2:1 remains valid if we consider the sum over both spin directions, but its separation into up and down spin contributions is nontrivial. Because the system is slightly above \(n=1\) filling (Fig.~\ref{fig.QPpeak_Jdep}) we also observe a spectral weight transfer to a band higher up in the spectrum, around \(\epsilon = 10\), that will be filled only for \(n>2\). The insets show the quasi-particle peak that we will discuss in Sec.~\ref{subsec.QPpeak}.

%%%%%%
\subsection{Double hopping term}
\label{subsec.doublehopping}
%%%%%%

\begin{figure}
\centering%
\includegraphics[width=1.0\linewidth]{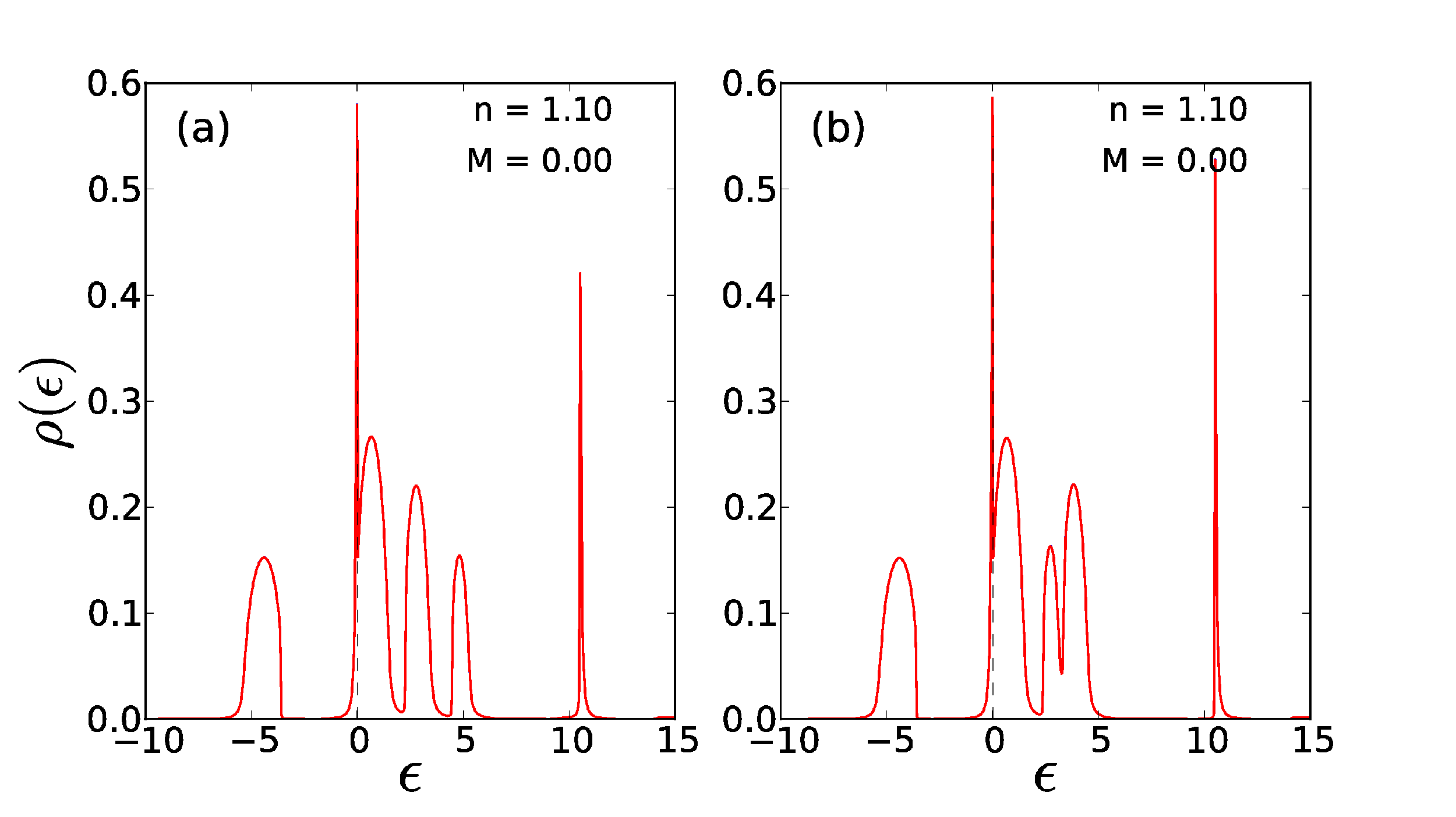}\\
\includegraphics[width=1.0\linewidth]{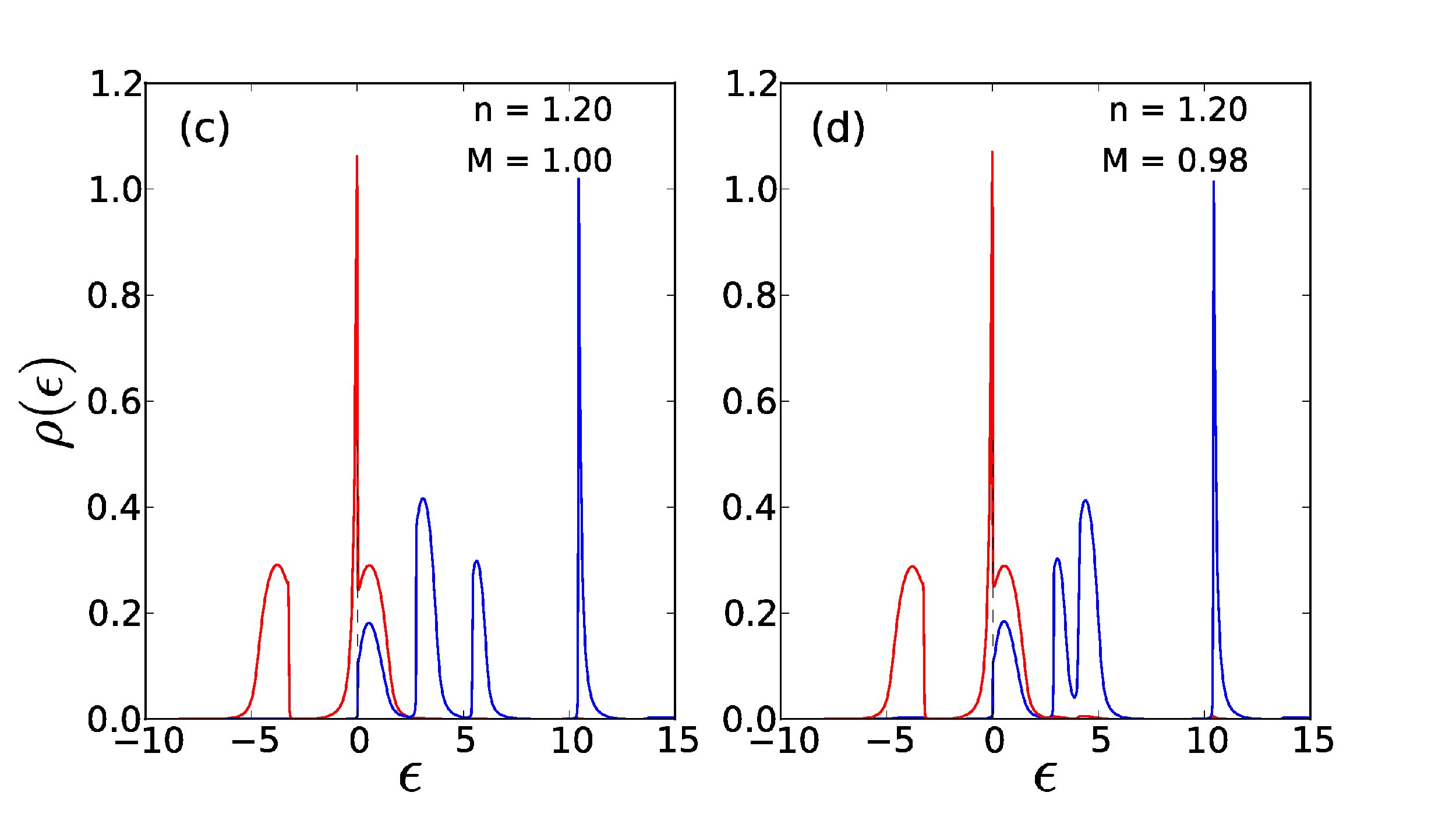}\\
\caption{\label{fig.Jp_Jp0}\(J^{\prime}=J\) (left) and \(J'=0\) (right) for \(J=1.0\) (top) and \(J=1.25\) (bottom). When the double hopping term is included one finds back the spectral weight as calculated by Lee and Philips, when it is not included the spectral weight of the upper Hubbard bands changes.}
\end{figure}

The pair hopping term is often neglected in studies investigating spectral functions at or around filling \(n=1\). This term is responsible for the splitting of a degenerate energy level (with energy \(U\)) at \(n=2\) into two different energy states \(U-J'\) and \(U+J'\). These energy states are rather high in the energy spectrum and for \( 0<n\lesssim1\) these bands are empty. In Fig.~\ref{fig.Jp_Jp0} \(n=1+x\), the Fermi energy lies on the edge of the first band of the series.
%FIGURE Double hopping term

The upper two graphs of Fig.~\ref{fig.Jp_Jp0} characterize non-magnetic states of the system. The double hopping term has influence only on the empty upper Hubbard bands, it does not change the magnetic phase of the system. In a system with the double hopping term we clearly find back the analytical results of Lee and Philips. Without this term the energy spectrum and the spectral weight in the upper spectrum change. It is the double hopping term that gives the spectral function its characteristic spectrum and spectral weight.

In the ferromagnetic case (lower part of Fig.~\ref{fig.Jp_Jp0}), we can see that the spectrum and spectral weight change in much the same way as in the paramagnetic case. The series of three consecutive upper Hubbard bands with the spectral weight distribution 3:2:1 can be obtained by summing the spin up and spin down spectral weights. The magnetization changes slightly when one leaves out the double hopping term, but overall the phase diagram remains the same with or without the double hopping term.

%%%%%%
\subsection{Quasi-particle peak}
\label{subsec.QPpeak}
%%%%%%
For doping levels around $n=1$ a quasi-particle peak is observed around the Fermi level (see inset of Fig.~\ref{fig.QPpeak_Jdep} and Fig.~\ref{fig.QPpeak_J125}). As our calculation gives \(\vec k\)-integrated Green functions, the small width of the peak \(W_{\mathrm{QP}}\) indicates a high effective mass \(m_{\mathrm{eff}} \sim 1/W_{\mathrm{QP}}\) of the corresponding quasi-particle. The peak is observed in the paramagnetic as well as in the ferromagnetic case. 
%FIGURE Quasi-particle peak
\begin{figure}
\centering%
\includegraphics[width=1.0\linewidth]{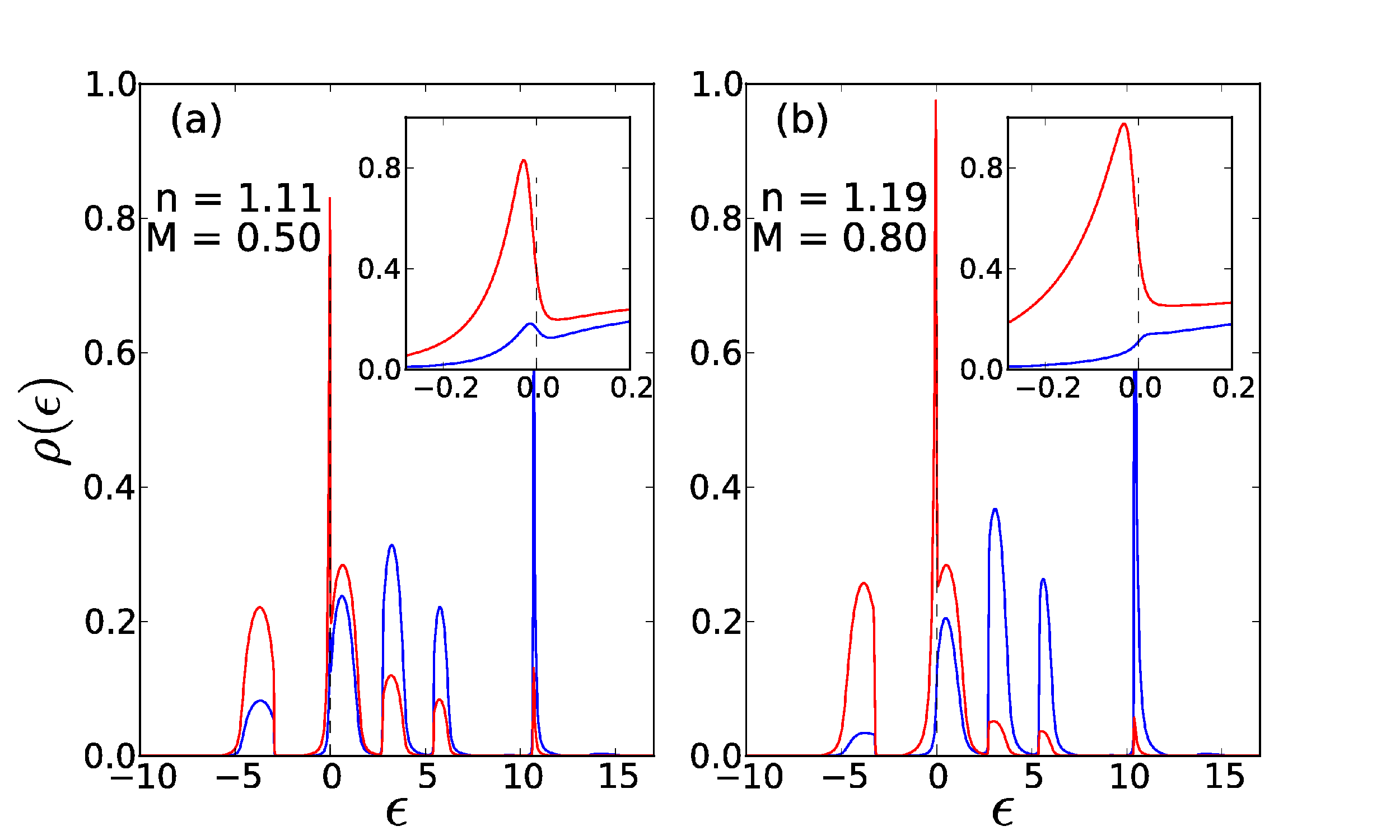}
\caption{\label{fig.QPpeak_J125}Quasi particle peak for \(n=1.11\) and \(n=1.19\), with \(J=1.25\) and \(T=4.3\times 10^{-3}\). The quasi-particle peak has the same spin as the majority spins of the lower Hubbard band.}
\end{figure}
On the right hand side of Fig.~\ref{fig.QPpeak_J125} the width of the 
quasi-particle peak is larger than on the left hand side. That is due to two effects: the increase of 
magnetization and of doping. In general, the peak width increases if we pass from the paramagnetic to 
the ferromagnetic state. 

The quasi-particle peak is strongly doping dependent. It exists around quarter filling ($n=1$) for electron and 
hole doping. The case of hole doping is shown in Fig.~\ref{fig.comparisonLee} (a). Exactly at $n=1$, we have an 
insulating state and there is no quasi-particle peak which is presented in Fig.~\ref{fig.QPpeak_J125} (a). We compare different doping values (no doping, 20 and 40 percent electron doping) and a tiny quasiparticle peaks on top 
of the main band exists only for low
doping values. The high effective mass is due to short range orbital correlations close to $n=1$ as will 
be explained more in detail in the next chapters. For $n=1.40$, the orbital correlations are weakened,
and the quasi-particle peak merges with the main band. Similar spectral function had already been calculated
before.\cite{Peters-2010fk} It is interesting to note that a tiny quasi-particle peak exists also close to half-filling
(see Fig.\ \ref{fig.comparisonLee} (b)). However, since we expect no tendency to orbital order at $n=2$, that
quasi-particle is probably of different origin and we did not investigate it further. 

%In Fig.~\ref{fig.comparisonLee} we also observe quasi-particle peaks at the Fermi energy. Figure~\ref{fig.comparisonLee} (a) concerns the hole-quasi particle equivalent of the (electronic) quasi-particle we will discuss in this article. The quasi-particle in Fig.~\ref{fig.comparisonLee} (b) can not be due to the mechanism we will discuss here, that is based on a ferromagnetic ground state. It is more likely to be supported by a mechanism with an anti-ferromagnetic ground state. 

%FIGURE Quasi-particle peak (ABSENCE)
\begin{figure}
\centering%
\includegraphics[width=1.0\linewidth]{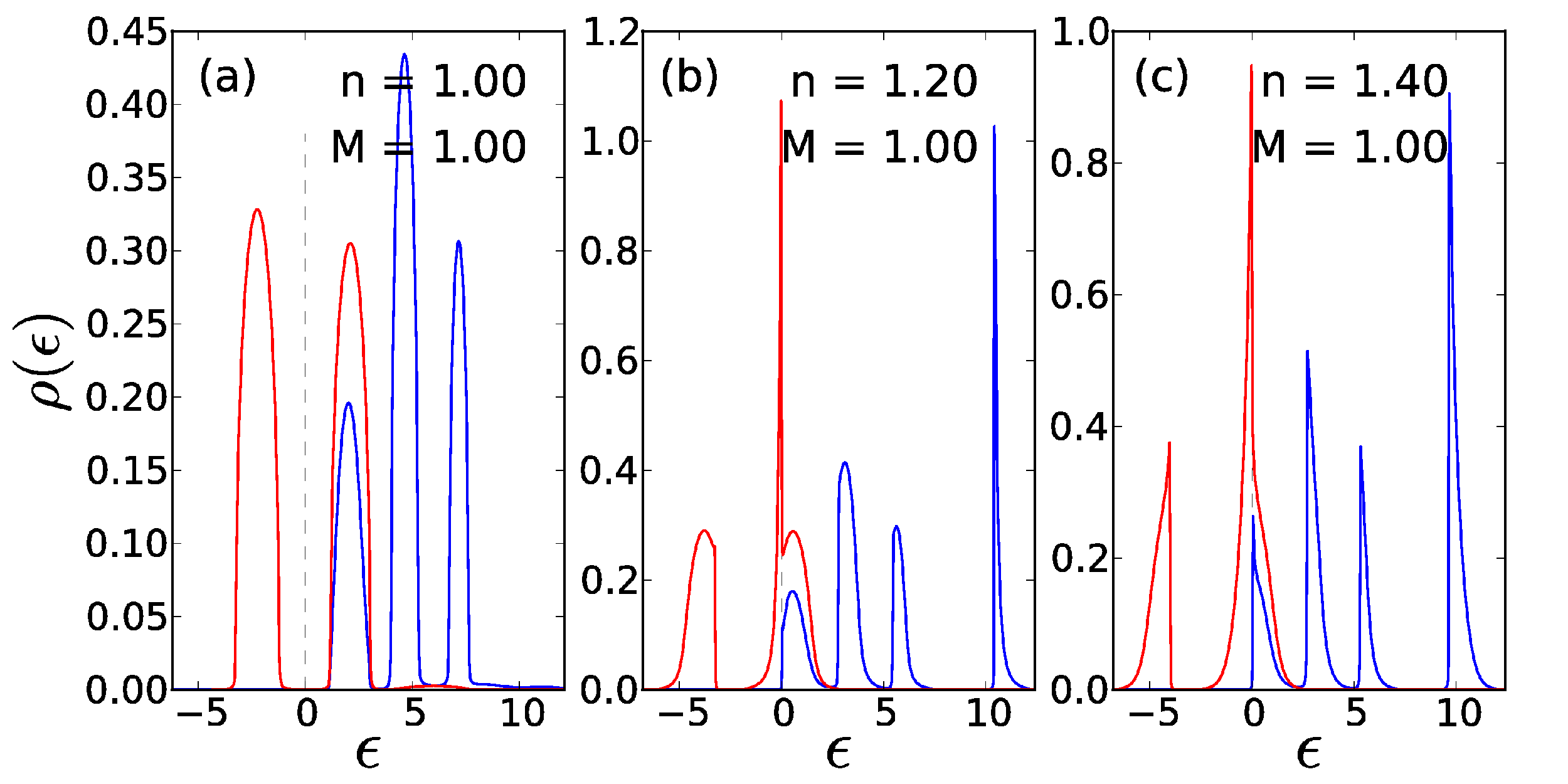}
\caption{\label{fig.QPpeak_J125}Appearance of the quasi particle peak for \(J=1.25\) and different doping values. 
The magnetization is very close to one in all three cases. 
The tiny quasi-particle peak is only visible for $n=1.20$, it is absent for \(n=1\), and it merges with the 
main band for high doping ($n=1.40$).} 
\end{figure}

In the ferromagnetic phase close to $n=1$ the quasi-particle is of the 
same spin as the majority spin of the lower Hubbard 
band. That is in strong contrast to the antiferromagnetic correlation between the local spin and the 
conduction electrons in the Kondo effect. Therefore, we have to find an alternative explanation that will be
presented in Section~\ref{sec.orbital_polaron}.

%%%%%%%%%%%%%%%%%%%%%
\section{Effective Hamiltonian for \(n=1\)}
\label{sec.Heff}
%%%%%%%%%%%%%%%%%%%%%

\begin{figure}
\centering%
\includegraphics[width=0.6\linewidth]{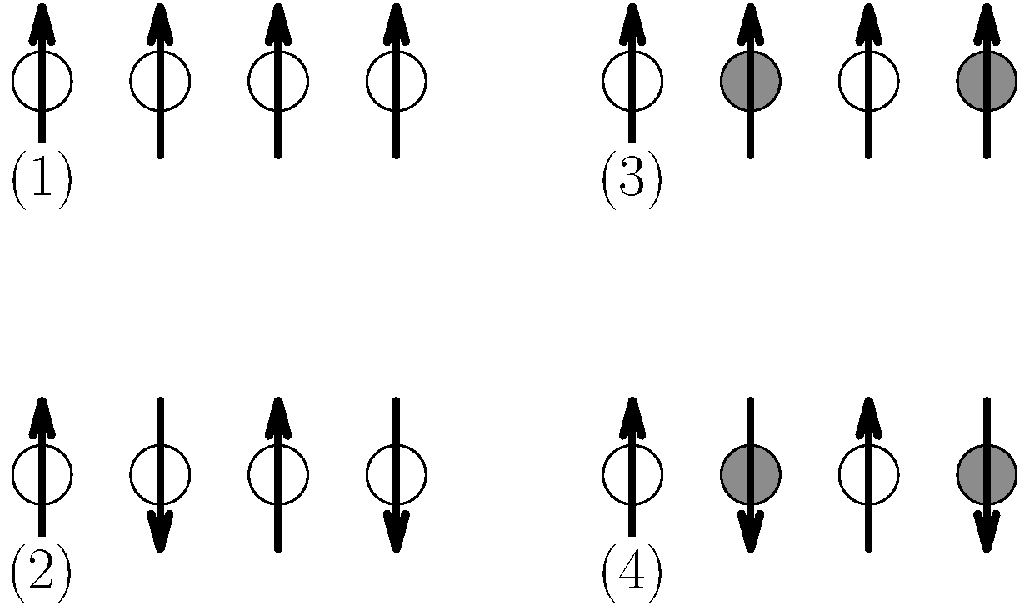}
\caption{\label{fig.possible_phases}The four possible phases of the system at $n=1$ are combinations of magnetic and orbital order.}
\end{figure}

To clarify the possible mechanism of quasi-particle formation one first has to analyze
the possible phases at $n=1$. There are in total four possible phases of the system, ferro- or 
antiferromagnetic and alternating or homogeneous orbital order, see Fig~\ref{fig.possible_phases}. 
The ferromagnetic and homogeneous orbital case is highest in energy since hopping between neighboring 
sites is not allowed. For this reason this configuration can be excluded. The phase could also be 
antiferromagnetic with homogeneous or alternating orbital order. In this case hopping is 
possible, but in the intermediate states the electrons meet at the same site with opposite spin and 
cannot benefit from Hund's exchange term. Lowest in energy is the ferromagnetic state with alternating 
orbital order. 
%FIGURE Possible phases

To be more precise we derive now an effective low-energy spin-pseudospin Hamiltonian \(\hat{H}_{\mathrm{LE}}\)
at $n=1$. It corresponds to the low-energy limit of the two-orbital Hubbard Hamiltonian (\ref{eq.Hkin})-(\ref{eq.HDE1}) assuming 
\begin{equation}
t\ll U,U^{\prime}, \ J<U \text{  and } \left\langle \hat{n}\right\rangle =1.\label{eq.tllU}
\end{equation}
Working in analogy to Refs.~\onlinecite{Kuei-1997fk},~\onlinecite{Oles-2005fk},~\onlinecite{Wohlfeld-2011fk}, we obtain the effective Hamiltonian in terms of the spin \(\hat{\bm{S}}\) and orbital pseudospin \(\hat{\bm{\tau}}\) operators (see Appendix \ref{app.Heff} for details):
\begin{align}
\hat{H}_{\mathrm{LE}} & = 
-\frac{t^{2}}{2}\sum_{R,g}\left\{ \frac{2}{U-3J}\left(\frac{1}{2}-2\hat{\bm{\tau}}_{R} \cdot \hat{\bm{\tau}}_{R+g} \right)  \right. \nonumber \\
& \times \left(\frac{3}{4}+\hat{\mathbf{S}}_{R}\hat{\mathbf{S}}_{R+g}\right) \nonumber \\
 & + \left[\frac{2}{U+J}\left(\frac{1}{4}+\hat{\tau}_{R}^{z}\hat{\tau}_{R+g}^{z}+\hat{\tau}_{R}^{x}\hat{\tau}_{R+g}^{x}-
 \hat{\tau}_{R}^{y}\hat{\tau}_{R+g}^{y}  \right)\right. \nonumber \\
 & +\left. \frac{4}{U-J}\left(\frac{1}{4}+\hat{\tau}_{R}^{y}\hat{\tau}_{R+g}^{y}\right)\right]
 \left.\left(\frac{1}{2}-2\hat{\mathbf{S}}_{R}\hat{\mathbf{S}}_{R+g}\right) \right\} \label{eq.Heffde} \,,
\end{align}
where we put \(U^{\prime} = U-2J\) and \(J^{\prime}=J\) to simplify the notation. This Hamiltonian corresponds to a special case \(\alpha=1\) (\(\alpha\) is the ratio between the transverse and longitudinal hopping terms) of the Hamiltonian of superexchange in the \(ab\) plane of alkali hyperoxide compounds RO\(_2\) (R=K, Kb, Cs), cf. Eqs.(1)-(3) of Ref.~\onlinecite{Wohlfeld-2011fk}.

It is not the aim of the present paper to analyze the low-energy Hamiltonian in detail. As it became 
clear from the qualitative considerations above, the ground state at \(n=1\) is the saturated ferromagnetic state with alternating orbital order. That is confirmed by the low-energy Hamiltonian. In the case of a saturated ferromagnetic system we have \( \hat{\bm{S}}_{R}\cdot \hat{\bm{S}}_{R+g} = \frac{1}{4}\). Then the low energy Hamiltonian (\ref{eq.Heffde}) reduces to the form
\begin{align}
\hat{H}_{\mathrm{LE}} & = \frac{2 t^2}{U-3J}  \sum_{R,g}  \hat{\bm{\tau}}_{R} \cdot \hat{\bm{\tau}}_{R+g} \,,\label{eq.reducedHLE1} \\
& = \mathcal{J}_0 \sum_{R,g}  \hat{\bm{\tau}}_{R} \cdot \hat{\bm{\tau}}_{R+g} \,,\label{eq.reducedHLE2}
\end{align}
that coincides with the Heisenberg Hamiltonian, but in orbital space. 
Because \(\mathcal{J}_0 = 2t^2/(U-3J)\) is positive, the minimal energy configuration will be the one with alternating orbital order. 
The system is thus in a ferromagnetic and alternating orbital phase.

%%%%%%%%%%%%%%%%%%%
\section{Orbital polaron}
\label{sec.orbital_polaron}
%%%%%%%%%%%%%%%%%%%

\begin{figure}[!t]
\centering%
\includegraphics[width=\linewidth]{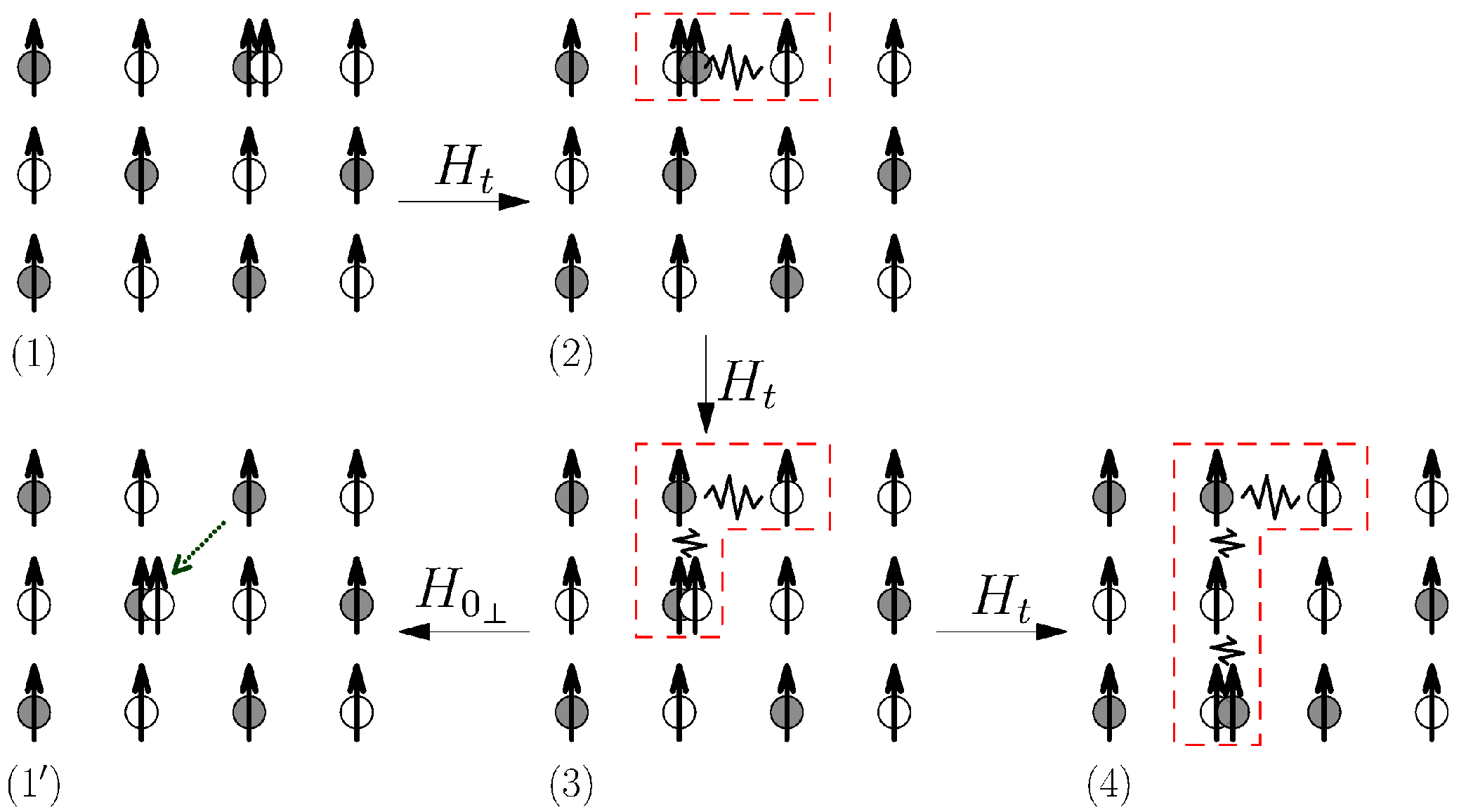}
\caption{\label{fig.orbital_polaron}Without orbital fluctuations the energy increases linearly with the length of the path traveled by the electron, whereas with orbital fluctuations the electron can move around without increasing the energy proportionally to the length of the path.}
\end{figure}

The here derived low-energy Hamiltonian also gives information about the mechanism that will be employed 
to make a charge carrier mobile on the lattice.
The charge carrier can be a hole or an electron and we illustrate in Fig.~\ref{fig.orbital_polaron} the case of electron doping (hole doping is equivalent). To describe the quasi-particle we have to add the kinetic energy term to the orbital Heisenberg Hamiltonian (\ref{eq.reducedHLE2}). 
In zeroth order approximation, this is the term (\ref{eq.Hkin})  with charge fluctuations projected out 
(cf. the discussion after Eq.(\ref{eq.Eij}) in Appendix \ref{app.Heff}) The resulting model is the very well known $t$-$J$ Hamiltonian. 
It was extensively studied in connection with high temperature superconductivity and it is quite interesting that our original model (\ref{eq.Hkin})-(\ref{eq.HDE1}) of double exchange ferromagnetism can be reduced to the same form. The only difference is that we have to replace spin fluctuations by orbital fluctuations. In the case of spin fluctuations, it is  known that the one-hole state in the antiferromagnetic background of the $t$-$J$ model gives rise to a coherent quasi-particle that is commonly called the spin polaron. Our case is very similar and leads to the orbital polaron. The orbital polaron was first proposed in the field of manganites. Taking a  modified $t$-$J$ model as the starting point, this kind of quasi-particle was shown to occur in the orbital-ordered ferromagnetic planes of LaMnO$_3$.\cite{Brink-2000kl} Since then it was studied in several other works.\cite{Daghofer-2004oq,Wohlfeld-2008ve,Wohlfeld-2009qf}
Here, we re-discover the orbital polaron in the complete double exchange Hamiltonian including
the double hopping term.

We are not going to repeat the studies of spectral properties in the framework of $t$-$J$ Hamiltonians. We just
remind here the basic arguments of the orbital polaron construction.
The electron does not move by simply jumping from nearest neighbor to nearest neighbor
since it would destroy the orbital order in such way (Fig.~\ref{fig.orbital_polaron}). To be precise we rewrite 
Eq.~(\ref{eq.reducedHLE2}) in a form that makes clear what the action of this low energy Hamiltonian is 
on the orbital pseudospin:
\begin{align}
\hat{H}_{\mathrm{LE}} &= H_{0_{\parallel}} + H_{0_{\perp}} \\
& = \mathcal{J}_0 \sum_{R,g} \hat{\tau}_{R}^{z}\hat{\tau}_{R+g}^{z} + \frac{\mathcal{J}_0}{2} \sum_{R,g} \left(\hat{\tau}_{R}^{+}\hat{\tau}_{R+g}^{-}+\hat{\tau}_{R}^{-}\hat{\tau}_{R+g}^{+}\right) \; . \nonumber
\end{align}
It is the second term which allows for orbital fluctuations. Without it, the electron would be localized,
since the application of $H_t$ leads to a string of misplaced orbitals with an energy increase that is 
proportional to the length of the path. It is the orbital fluctuation term
$H_{0\perp}$ which may reduce the length of the path. In such a way, it delocalizes the electron and we therefore expect a quasi-particle width of the order of $\mathcal{J}_0$.
%FIGURE Orbital polaron

The band-width of the spin polaron is known for the 2D $t$-$J$ model to be approximately 2$J_H$ where
$J_H$ is the exchange constant of the Heisenberg term. Our study is different in many respects. For instance, the 
semi-circular DOS corresponds to a Bethe lattice instead of a 2D quadratic one. Also, it is not sure whether 
the single site DMFT is able to reproduce the correct quasi-particle band width. Nevertheless, it can be 
expected that the quasi-particle band width should be proportional to  $\mathcal{J}_0$ close
to $n=1$. That is confirmed by the half-width at half-maximum of the quasi particle peak $W_{QP}$ which can be
read out of Fig.~(\ref{fig.QPpeak_J125}) to be $W_{QP} \approx 0.12$ in a state which is 
close to saturated FM but still has low doping. Therefore,  $W_{QP}$ is approximately equal to $\mathcal{J}_0$ which seems to be a realistic value. But one probably needs a more detailed method such as cluster DMFT to observe the correct dependencies of the quasi-particle bandwidth on the parameter values of the model for a given lattice type.

%%%%%%%%%%%%%%%%%%%
\section{Discussion and Conclusion}
\label{sec.conclusion}
%%%%%%%%%%%%%%%%%%%

We investigated the complete two-orbital Hubbard model including the double-hopping term by means of the 
single site dynamical mean field theory (DMFT). We found ferromagnetism in the electron doped case $n>1$
that may extend up to $n=1$ for sufficiently high values of the Hund exchange coupling $J$. The spectral 
function has two very characteristic features, a narrow quasi-particle peak at the Fermi level with a spin 
parallel to the majority spin direction and a characteristic equidistant three-peak structure with a 
weight distribution 3:2:1. We found the double-hopping term to be of crucial importance for obtaining the 
correct spectral function.

Our study unifies several previous works. The magnetic phase diagram of the present model, but without
double hopping term, was already published before.\cite{Peters-2010fk,Peters-2011uq} 
Our numerical results of the characteristic three-peak structure confirm the analytical predictions of Lee and Philips for the complete model
including the double hopping term.\cite{Lee-2011uq}  We interpret the narrow quasi-particle peak as an orbital polaron that
was known before in the field of manganites.\cite{Brink-2000kl} These former studies considered the one-hole 
problem in the effective, low-energy, $t$-$J$ like Hamiltonian. The present study is less restrictive since it 
treats the full two-orbital Hubbard model that is valid close to the Fermi level as well as for the 
high-energy features. Furthermore, it works for arbitrary doping.

We admit a weakness of our approach since we did not calculate the two-sublattice problem. That would be necessary to obtain the correct alternating orbital-ordered phase at and close to $n=1$ and for large $U$. It was obtained by Pruschke and Peters \cite{Peters-2010fk} by 
DMFT and we confirm it here by the derivation of an effective spin-pseudospin Hamiltonian for $n=1$. We have shown that this 
Hamiltonian leads to a ferromagnetic ground state with alternating orbital order. Adding additional charge 
carriers, we end up with the same $t$-$J$ Hamiltonian as has been known for cuprate superconductors for a long time. This way we can interpret the observed quasi-particle as an orbital polaron in close analogy to the spin polaron. The charge carriers have the tendency to destroy the orbital order (see Fig.~\ref{fig.orbital_polaron}) and we expect the orbitally ordered phase in a small region around $n=1$ only. That is the reason why we restricted our study to the orbital liquid phase where, however, short range alternating orbital correlations persist. These correlations are visible in the spectral function by the small bandwidth of the quasi-particle. We have shown that the quasi-particle peak broadens with increased doping indicating in such a way the decrease of orbital correlations.    

The two-orbital Hubbard Hamiltonian is relevant for several material classes. First of all it concerns 
manganites like LaMnO$_3$ and related compounds with an electron configuration close to  
\(t_{2g}^3 \ e_g^1\) . In that case, besides a partially filled $e_g$ shell that is well described by our model,
an additional local spin ($S=3/2$) exists. That additional spin may lead to modifications that are outside the scope of the present model. It is better suited for nickelates like doped or undoped  LiNiO$_2$, NaNiO$_2$ or  AgNiO$_2$, with an electron configuration close to \(t_{2g}^6 \ e_g^1\) (completely filled \(t_{2g}\) shells).\cite{Dare-2003uq,Vernay-2004kx,Sugiyama-2010uq} In undoped LiNiO\(_2\) the e\(_g\) orbitals are degenerate and there is no orbital or magnetic order in that specific two-dimensional triangular lattice (see Refs. \onlinecite{Dare-2003uq} and \onlinecite{Vernay-2004kx} and references therein). However, electron doping leads to a ferromagnetic phase as it was observed recently\cite{Sugiyama-2010uq} and it reminds of our phase diagram (Fig.~\ref{fig.phasediag_J125}). Manganites and nickelates are just two examples with partially filled $e_g$ shells but there exist many more possibilities in the field of transition metal compounds (like LaVO$_3$ for example).\cite{Wohlfeld-2009qf} As it was already mentioned in the introduction, the double exchange mechanism was also discussed for diluted magnetic semiconductors 
(DMS).\cite{Krstajic-2004kx} In many cases, the double exchange mechanism is in competition with other 
mechanisms for ferromagnetism as for example itinerant magnetism, direct exchange, Zener's kinetic $p$-$d$ exchange, 
and so on. Our results for the spectral function provide a unique means to distinguish it from other 
mechanisms. This concerns especially the quasi-particle peak at the Fermi level. If, in a given ferromagnetic
material a heavy quasi-particle peak with a spin parallel to the majority spin direction is observed, it
would give a strong hint in favor of the double exchange mechanism. To observe it, we propose spin resolved
photoemission with polarized photons.\cite{Osterwalder-2006uq} Another characteristic feature is the three-peak
structure with weight distribution 3:2:1 in the unoccupied part of the spectral function. 
Since the two distances between the three peaks are found to be equal, and very close to \(2 J\), it gives an experimental means to determine directly the Hund exchange coupling~$J$.

To interpret the quasi-particle peak, we concentrated the discussion on orbital fluctuations in a saturated 
ferromagnetic state. But our DMFT simulation is also able to describe spin fluctuations. These are present in the paramagnetic case. One might expect that the coupling to spin fluctuations makes the quasi-particle
even more heavy. And indeed, we observed a reduction of the quasi-particle band width if the system
switches from the ferromagnetic to the paramagnetic phase. But a detailed theory of that process has still to be
developed. The spin-pseudospin Hamiltonian which was derived in the present work may serve as a
starting point for such a theory.\\

%%%%%%%%%%%%%
\appendix
%%%%%%%%%%%%%

%%%
\section{Effective Hamiltonian}
\label{app.Heff}
%%%%%
Here we give the details of the derivation of the effective low-energy Hamiltonian (\ref{eq.Heffde}). 
We write 
\begin{equation}
\hat{H}_{d} = \sum_{i}\left|i\right\rangle E_{i}\left\langle i\right|,\ \hat{H}_t=\sum_{i\neq j}\left|i\right\rangle t_{ij}\left\langle j\right|,
\end{equation}
where the eigenvalues \(E_{i}\) and eigenvectors \(\left|i\right\rangle \)
of the unperturbed Hamiltonian \(\hat{H}_{d}\) are assumed to be known.
Without loss of generality, we may assume that the perturbation \(\hat{H}_t\)
contains only non-diagonal terms. The operator
\begin{equation}
\label{eq.W}
\hat{W}=\sum_{i\neq j}\left|i\right\rangle \frac{t_{ij}}{E_{j}-E_{i}}\left\langle j\right| ,
\end{equation}
has the property 
\begin{equation}
\left[\hat{H}_{d},\hat{W}\right]\equiv\hat{H}_{d}\hat{W}-\hat{W}\hat{H}_{d}=-\hat{H}_t.\label{eq.H0W}
\end{equation}
Then, up to second order a canonical transformation gives an effective Hamiltonian
\begin{align}
\label{eq.Heff}
\hat{H}^{\mathrm{eff}} & =  \exp(-\hat{W})\hat{H}\exp(\hat{W})=\hat{H}_{d}+\hat{H}_{\mathrm{LE}}   \\
 & =  \hat{H}_{d}+\frac{1}{2}\left[\hat{H}_t,\hat{W}\right].\nonumber 
\end{align}
 An explicit calculation gives 
\begin{equation}
\label{eq.HLE}
\hat{H}_{\mathrm{LE}}  =  \frac{1}{2}\sum_{i,f,j}\left|f\right\rangle t_{fj}t_{ji}D_{fji}\left\langle i\right|,
\end{equation}
with
\begin{align}
D_{fji} & \equiv  \left(\frac{1}{E_{fj}}-\frac{1}{E_{ji}}\right),                      \label{eq.Dnjm}\\
E_{ij} & \equiv  E_{i}-E_{j}.        \label{eq.Eij}
\end{align}
It is clear from Eqs.~(\ref{eq.HLE})-(\ref{eq.Eij}) that the transformation may eliminate via Eq.~(\ref{eq.H0W})  only those terms in
\(\hat{H}_t\) that connect the states with different energy, and it makes sense only when \(|t_{ij}| \ll E_{ij}\). 
For our system this means that we eliminate only hoppings between the states which differ by total numbers of 
\(n\)-occupied sites, \(n=0,1,2,3,4\). At quarter filling 
the states \(\left|i\right\rangle\) and  \(\left|f\right\rangle\) in Eq.~(\ref{eq.HLE})
contain one particle at each site, and in the intermediate states
\(\left|j\right\rangle \) one site is empty and a neighboring site
is doubly occupied. The Hamiltonian \(\hat{H}_{\mathrm{LE}}\) in Eq.~(\ref{eq.HLE})
will contain the interactions between neighboring sites.

In the Table I of Ref.~\onlinecite{Lee-2011uq} a complete list of the eigenstates
and eigenenergies of the Hamiltonian \(\hat{H}_{d}\) (\ref{eq.HDE1})-(\ref{eq.Hd})
is given. We shall denote them as following: \(\left|0\right\rangle \) is the vacuum state, number of electrons
\(n=0\), energy \(E_{0}=0\). \(d_{r,\alpha,\sigma}^{\dagger}\left|0\right\rangle\) is a one-particle
state \(n=1\), with \(E_{\alpha,\sigma}=0\); then, e.g., 
\begin{equation}
\label{eq.i}
\left|i\right\rangle =\prod_{r,\alpha_{r},\sigma_{r}}d_{r,\alpha_{r},\sigma_{r}}^{\dagger}\left|0\right\rangle .
\end{equation}
We will characterize two-particle states at lattice site \(r\) by their energy and,
when needed, the total spin projection
\begin{align}
&\left|r,U^{\prime}-J,2\sigma\right\rangle =  d_{r,a,\sigma}^{\dagger}d_{r,b,\sigma}^{\dagger}\left|0\right\rangle ,\nonumber \\
&\left|r,U^{\prime}-J,0\right\rangle  =  \frac{1}{\sqrt{2}}\left(d_{r,a,\uparrow}^{\dagger}d_{r,b,\downarrow}^{\dagger}+d_{r,a,\downarrow}^{\dagger}d_{r,b,\uparrow}^{\dagger}\right)\left|0\right\rangle ,\nonumber \\
&\left|r,U^{\prime}+J\right\rangle  =  \frac{1}{\sqrt{2}}\left(d_{r,a,\uparrow}^{\dagger}d_{r,b,\downarrow}^{\dagger}-d_{r,a,\downarrow}^{\dagger}d_{r,b,\uparrow}^{\dagger}\right)\left|0\right\rangle ,\nonumber \\
&\left|r,U-J^{\prime}\right\rangle =  \frac{1}{\sqrt{2}}\left(d_{r,a,\uparrow}^{\dagger}d_{r,a,\downarrow}^{\dagger}-d_{r,b,\uparrow}^{\dagger}d_{r,b,\downarrow}^{\dagger}\right)\left|0\right\rangle ,\nonumber \\
&\left|r,U+J^{\prime}\right\rangle =  \frac{1}{\sqrt{2}}\left(d_{r,a,\uparrow}^{\dagger}d_{r,a,\downarrow}^{\dagger}+d_{r,b,\uparrow}^{\dagger}d_{r,b,\downarrow}^{\dagger}\right)\left|0\right\rangle \nonumber .
\end{align}

%%%
The effective second-order Hamiltonian (\ref{eq.HLE}) has the same
form for every particular bond (a couple of neighboring sites) \(R\),
\(R^{\prime}=R+g\) . So, we derive it for a two-site system, and
the total Hamiltonian will be the sum over all bonds. The action of
the operator \(\hat{H}_t\) on \(\left|i\right\rangle\) (\ref{eq.i})
is 
\begin{widetext}
\begin{align}
 \hat{H}_t d_{R,\alpha,\sigma}^{\dagger}d_{R^{\prime},\alpha^{\prime},\sigma^{\prime}}^{\dagger}\left|0\right\rangle &= 
 -t\Big\{ \delta_{\sigma^{\prime},\sigma}\delta_{\alpha^{\prime},-\alpha}d_{R,-\alpha,\sigma}^{\dagger}d_{R,\alpha,\sigma}^{\dagger}\Big.
  +\delta_{\sigma^{\prime},-\sigma}\left(\delta_{\alpha^{\prime},\alpha}d_{R,\alpha,-\sigma}^{\dagger}d_{R,\alpha,\sigma}^{\dagger}
  +\left.\delta_{\alpha^{\prime},-\alpha}d_{R,-\alpha,-\sigma}^{\dagger}d_{R,\alpha,\sigma}^{\dagger}\right)\right. \\
 & \qquad +\Big.\left(R\leftrightarrow R^{\prime}\right)\Big\} \left|0\right\rangle \nonumber \\ 
 &=t \Big\{ \delta_{\sigma^{\prime},\sigma}\delta_{\alpha^{\prime},-\alpha}\left(\delta_{\alpha a}-\delta_{\alpha b}\right)\left|R,U^{\prime}-J,2\sigma\right\rangle \Big. +\delta_{\sigma^{\prime},-\sigma}\frac{1}{\sqrt{2}}\left[\delta_{\alpha^{\prime},\alpha}\left(\delta_{\sigma\uparrow}-\delta_{\sigma\downarrow}\right)\left(\delta_{\alpha a}+\delta_{\alpha b}\right)\left|R,U+J^{\prime}\right\rangle \right.\nonumber \\
& \qquad+\delta_{\alpha^{\prime},\alpha}\left(\delta_{\sigma\uparrow}-\delta_{\sigma\downarrow}\right)\left(\delta_{\alpha a}-\delta_{\alpha b}\right)\left|R,U-J^{\prime}\right\rangle 
  +\delta_{\alpha^{\prime},-\alpha}\left(\delta_{\sigma\uparrow}-\delta_{\sigma\downarrow}\right)\left(\delta_{\alpha a}+\delta_{\alpha b}\right)\left|R,U^{\prime}+J\right\rangle \nonumber \\
& \qquad \left.+\delta_{\alpha^{\prime},-\alpha}\left(\delta_{\sigma\uparrow}+\delta_{\sigma\downarrow}\right)\left(\delta_{\alpha a}-\delta_{\alpha b}\right)\left|R,U^{\prime}-J,0\right\rangle \right] \Big. +\left(R\leftrightarrow R^{\prime}\right)\Big\} \nonumber.
 \end{align}
 \end{widetext}
 This expression provides us the matrix elements, which enter \(\hat{H}_{\mathrm{LE}}\)
(\ref{eq.HLE}). Now, we are ready to write the effective Hamiltonian
for one bond and at \(n=1\).

\begin{widetext}
\begin{eqnarray}
\hat{H}_{\mathrm{LE}} & = & \sum_{\alpha,\alpha^{\prime},\beta,\beta^{\prime}}\sum_{\sigma,\sigma^{\prime},s,s^{\prime}}\sum_{j}d_{R,\alpha,\sigma}^{\dagger}d_{R^{\prime},\alpha^{\prime},\sigma^{\prime}}^{\dagger}\left\langle 0\right|d_{R^{\prime},\alpha^{\prime},\sigma^{\prime}}d_{R,\alpha,\sigma}\hat{H}_t\left|j\right\rangle \left\langle j\right|\hat{H}_td_{R,\beta,s}^{\dagger}d_{R^{\prime},\beta^{\prime},s^{\prime}}^{\dagger}\left|0\right\rangle \frac{1}{0-E_{j}}d_{R^{\prime},\beta^{\prime},s^{\prime}}d_{R,\beta,s} \nonumber \\
 & = & -2t^{2}\sum_{\alpha,\beta,\sigma,s}\left\{ \frac{1}{U^{\prime}-J}d_{R,\alpha,\sigma}^{\dagger}d_{R^{\prime},-\alpha,\sigma}^{\dagger}d_{R^{\prime},-\beta,s}d_{R,\beta,s}\delta_{\sigma s}\left(\delta_{\alpha a}-\delta_{\alpha b}\right)\left(\delta_{\beta a}-\delta_{\beta b}\right)\right.\nonumber \\
 & + & \frac{1}{2}\left[\frac{1}{U+J^{\prime}}d_{R,\alpha,\sigma}^{\dagger}d_{R^{\prime},\alpha,-\sigma}^{\dagger}d_{R^{\prime},\beta,-s}d_{R,\beta,s}\left(\delta_{\sigma\uparrow}-\delta_{\sigma\downarrow}\right)\left(\delta_{s\uparrow}-\delta_{s\downarrow}\right)\right.\nonumber \\
 & + & \frac{1}{U-J^{\prime}}d_{R,\alpha,\sigma}^{\dagger}d_{R^{\prime},\alpha,-\sigma}^{\dagger}d_{R^{\prime},\beta,-s}d_{R,\beta,s}\left(\delta_{\alpha a}-\delta_{\alpha b}\right)\left(\delta_{\beta a}-\delta_{\beta b}\right)\left(\delta_{\sigma\uparrow}-\delta_{\sigma\downarrow}\right)\left(\delta_{s\uparrow}-\delta_{s\downarrow}\right)\nonumber \\
 & + & \frac{1}{U^{\prime}+J}d_{R,\alpha,\sigma}^{\dagger}d_{R^{\prime},-\alpha,-\sigma}^{\dagger}d_{R^{\prime},-\beta,-s}d_{R,\beta,s}\left(\delta_{\sigma\uparrow}-\delta_{\sigma\downarrow}\right)\left(\delta_{s\uparrow}-\delta_{s\downarrow}\right)\nonumber \\
 & + & \left.\left.\frac{1}{U^{\prime}-J}d_{R,\alpha,\sigma}^{\dagger}d_{R^{\prime},-\alpha,-\sigma}^{\dagger}d_{R^{\prime},-\beta,-s}d_{R,\beta,s}\left(\delta_{\alpha a}-\delta_{\alpha b}\right)\left(\delta_{\beta a}-\delta_{\beta b}\right)\right]\right\} \label{eq.HLE2}
 \end{eqnarray}
 \end{widetext}
 Note that intermediate doubly occupied states \(\left|j\right\rangle\)
may be located on \(R\) as well as on \(R^{\prime}\) sites. This gives
a factor 2 in the Eq.(\ref{eq.HLE2}).

When \(n=1\), we may express the product of operators at the same site
\(d_{R,\alpha,\sigma}^{\dagger}d_{R,\beta,s}\) , depending on the indices
\(\alpha,\beta\) and \(\sigma,s\), via the product of spin \(\hat{S}\)
and pseudospin \(\hat{\tau}\) operators, according to the rules (see
Eq. (3) of the Ref.~\onlinecite{Kugel-1973fl})
\begin{align}
\label{eq.KKrules}
(\alpha,\beta = \alpha)\rightarrow1/2+\alpha\hat{\tau}^{z}, \quad &(\sigma,s =  \sigma)\rightarrow1/2+\sigma\hat{S}^{z},\nonumber \\
 (a,b)\rightarrow\hat{\tau}^{+}, \qquad &(\uparrow,\downarrow)\rightarrow\hat{S}^{+}, \\
(b,a)\rightarrow\hat{\tau}^{-}, \qquad &(\downarrow,\uparrow)\rightarrow\hat{S}^{-}. \nonumber
\end{align}
Thus, e.g., $d_{R,a,\uparrow}^{\dagger}d_{R,b,\uparrow}=\hat{\tau}_{R}^{+}\left(1/2+\hat{S}_{R}^{z}\right)$,
etc. All terms in Eq.~(\ref{eq.HLE2}), should be substituted according to the rules in Eq.~(\ref{eq.KKrules}). For the term \(\propto-t^{2}/(U+J^{\prime})\) for example we get
\begin{widetext}
\begin{align}
 \sum_{\alpha,\beta,\sigma,s}&d_{R,\alpha,\sigma}^{\dagger}d_{R^{\prime},\alpha,-\sigma}^{\dagger}d_{R^{\prime},\beta,-s}d_{R,\beta,s}\left(\delta_{\sigma\uparrow}-\delta_{\sigma\downarrow}\right)\left(\delta_{s\uparrow}-\delta_{s\downarrow}\right)=\nonumber \\
 & = \sum_{\alpha,\beta}\left(d_{R,\alpha,\uparrow}^{\dagger}d_{R^{\prime},\alpha,\downarrow}^{\dagger}-d_{R,\alpha,\downarrow}^{\dagger}d_{R^{\prime},\alpha,\uparrow}^{\dagger}\right)\left(d_{R^{\prime},\beta,\downarrow}d_{R,\beta,\uparrow}-d_{R^{\prime},\beta,\uparrow}d_{R,\beta,\downarrow}\right)\nonumber \\
 & = \left(\frac{1}{2}+2\hat{\tau}_{R}^{z}\hat{\tau}_{R^{\prime}}^{z}+\hat{\tau}_{R}^{+}\hat{\tau}_{R^{\prime}}^{+}+\hat{\tau}_{R}^{-}\hat{\tau}_{R^{\prime}}^{-}\right)\left(\frac{1}{2}-2\hat{\mathbf{S}}_{R}\hat{\mathbf{S}}_{R^{\prime}}\right)\label{eq.UplusJp}
\end{align}
\end{widetext}
After this transformation we obtain the effective Hamiltonian 
\begin{widetext}
\begin{align}
\hat{H}_{\mathrm{LE}} & = -\frac{t^{2}}{2}\sum_{R,g}\left\{ \frac{2}{U^{\prime}-J}\left(\frac{1}{2}-2\hat{\bm{\tau}}_{R} \cdot \hat{\bm{\tau}}_{R+g} \right)   \left(\frac{3}{4}+\hat{\mathbf{S}}_{R}\hat{\mathbf{S}}_{R+g}\right) \right. \nonumber \\
 & + \left[\frac{1}{U+J^{\prime}}\left(\frac{1}{2}+2\hat{\tau}_{R}^{z}\hat{\tau}_{R+g}^{z}+\hat{\tau}_{R}^{+}\hat{\tau}_{R+g}^{+}+\hat{\tau}_{R}^{-}\hat{\tau}_{R+g}^{-}  \right)\right.\nonumber \\
 & + \frac{1}{U-J^{\prime}}\left(\frac{1}{2}+2\hat{\tau}_{R}^{z}\hat{\tau}_{R+g}^{z}-\hat{\tau}_{R}^{+}\hat{\tau}_{R+g}^{+}-\hat{\tau}_{R}^{-}\hat{\tau}_{R+g}^{-}\right)\nonumber \\
 & + \left.\frac{1}{U^{\prime}+J}\left(\frac{1}{2}-2\hat{\tau}_{R}^{z}\hat{\tau}_{R+g}^{z}+\hat{\tau}_{R}^{+}\hat{\tau}_{R+g}^{-}+\hat{\tau}_{R}^{-}\hat{\tau}_{R+g}^{+}\right)\right] \left.\left(\frac{1}{2}-2\hat{\mathbf{S}}_{R}\hat{\mathbf{S}}_{R+g}\right)\right\} \label{eq.Heffde2} \; .
\end{align}
\end{widetext}
Recalling that \(J^{\prime}=J, U^{\prime}=U-2J \), and \(\tau^{\pm} = \tau^x \pm \I\tau^y \) we obtain Eq. (\ref{eq.Heffde}).

\end{document}